\documentclass[aps,prl,groupedaddress,twocolumn, 
a4paper, longbibliography, 
]{revtex4-2}
\usepackage{amsmath}
\usepackage[english]{babel}
\usepackage{hyperref}

\usepackage[]{natbib}

\usepackage{graphicx}
\usepackage{siunitx}
\newcommand\ket[2][]{\ensuremath{#1\lvert {#2} #1\rangle}}

\begin{document}

\title{Quectonewton local force sensor}

\author{Yann Balland}
\author{Luc Absil}
\author{Franck Pereira Dos Santos}
\affiliation{LNE-SYRTE}

\date{\today}

\begin{abstract}
  We report on the realization of a quantum sensor based on trapped atom interferometry in an
  optical lattice for the measurement of atom-surface interactions, with sub-micrometer-level
  control of the mean atom-surface separation distance. The force sensor reaches a short-term
  sensitivity of \qty{3.4e-28}{N} at \qty{1}{s} and a long-term stability of 4 qN
  (\qty{4e-30}{N}). We perform force measurements in the 0--\qty{300}{\um} range, and despite significant stray forces caused by adsorbed atoms on the surface, we obtain evidence of the Casimir-Polder force.
\end{abstract}

\maketitle

Short-range forces are one of the many frontiers of modern physics
\cite{wolfOpticalLatticeClocks2007,Antoniadis2011}.
In the submillimeter scales, quantum electrodynamics (QED) interactions are dominant, and give rise
in the case of atom-surface interactions to the Casimir-Polder force \cite{PhysRev.73.360}. Since
the first highlight of this force \cite{raskinInteractionNeutralAtomic1969}, several
different methods \cite{laliotis_atom-surface_2021} have been able to bring out Casimir-Polder
forces, notably by measuring the transmission of an atomic beam through a micron-sized cavity
\cite{Sukenik1993}, diffracting matter waves on a surface \cite{garcion_intermediate-range_2021} or
performing spectroscopy in vapor cells \cite{blochAtomwallInteraction2005, peyrot_measurement_2019}. 
However these approaches have struggled to achieve the high measurement sensitivity required to detect the very weak forces involved all while maintaining a good understanding of the setup geometry, particularly the distance separating atoms from the surface.

Few experiments have achieved measuring Casimir-Polder forces while controlling directly the atom-surface distance. In the range from tens to hundreds of nanometers, the Casimir-Polder potential was measured directly by reflecting the atoms on an evanescent field \cite{landraginMeasurementVanWaals1996, bender_direct_2010}. In the micrometer range (around
\qty{6}{\um}), the Casimir-Polder effect has been highlighted by tracking the oscillations of a Bose-Einstein Condensate (BEC) inside a harmonic trap~\cite{harber_measurement_2005}, a method that does not measure the force though, but rather its gradient. 
An appealing method that we will use here to perform spatially resolved and sensitive force measurements is to trap atoms during extended measurement times in a shallow vertical optical lattice.

\begin{figure}
  \centering
  \includegraphics[]{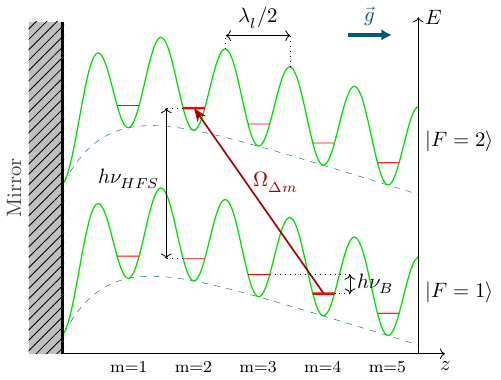}
  \caption{Wannier-Stark ladder potential created by the optical lattice retroreflected on a mirror. Two neighbouring states $\ket{W_m, F=1}$ and $\ket{W_{m+\Delta m}, F=2}$ separated by the energy $h(\nu_{HFS} + \Delta m
  \,\nu_B)$ can be coupled by Raman transitions with a Rabi frequency $\Omega_{\Delta m}$.
  Close to the surface, Casimir-Polder interactions modify the external potential (dashed lines) and
  perturb the value of $\nu_B$.
 }\label{fig:WS_ladder}
\end{figure}

The lattice potential, tilted by gravity $g$ or any other external uniform force $F_e$, leads to a
Wannier-Stark Hamiltonian. Its eigenstates $\ket{W_m}$ are localized in every well $m$ of the
lattice and separated by the Bloch frequency $\nu_b = (m_\mathrm{Rb}\, g + F_e) \lambda_l / (2 h)$, where
$m_\mathrm{Rb}$ is the mass of the atom and $\lambda_l$ the wavelength of the lattice
beam~\cite{pelissonLifetimesAtomsTrapped2013} (Fig. \ref{fig:WS_ladder}). The Wannier-Stark ladder
holds even if the $F_e(z)$ force is not uniform but remains perturbative relatively to the gravity
force $m_\mathrm{Rb}\, g$. Different protocols have been proposed to measure Casimir-Polder forces using such optical lattices, based for instance on performing interferometry with a BEC \cite{Dimopoulos2003} or measuring the frequency shift $\delta \nu_b$ of Bloch oscillations \cite{Carusotto2005,sorrentino_quantum_2009}. 
In the experiment reported here, which shares common features with these last two methods, we use Raman transitions to couple two eigenstates
distant of $\Delta m$ wells apart~\cite{beaufils_laser_2011},
as illustrated in Fig. \ref{fig:WS_ladder}. Using Rabi spectroscopy or Ramsey interferometry, we
then probe the corresponding transition energy $\Delta m \,\nu_b$ with the very high
sensitivity associated to frequency measurements with an atomic clock, and thus the external force applied to the atoms.

We present here the features and performances of such a high sensitivity quantum sensor for local
force measurements. We demonstrate a short-term sensitivity of \qty{3.4e-28}{N} at \qty{1}{s} and a
long-term stability of \qty{4e-30}{N} on the force measurements, which allows us to measure
interaction forces at the percent level in the range of tens of micrometers. Finally, we compare our measurements with the expected Casimir-Polder force.

$\mathrm{^{87}Rb}$ atoms are first laser cooled in a magneto-optical trap, before being transferred inside a crossed optical dipole trap. A \qty{2}{s} long evaporative cooling stage leaves us with about \num{120000}
atoms at a temperature of \qty{300}{nK}. The vertical size of the cloud is estimated to an rms radius
of \qty{10}{\um}. By adiabatically ramping up the dipolar trap at
the end of the evaporation, we reduce this size by a factor of 3, at the cost of heating the atoms
up to \qty{1500}{nK}. This preparation step is done in a first chamber, \qty{30}{cm} below our surface of  interest, a silica dielectric dichroic mirror, highly reflective at \qty{532}{nm} for our measurement shallow lattice, but transparent at \num{780} and \qty{1064}{nm}, the wavelengths of our other lasers. 

\begin{figure}
  \centering
  \includegraphics[]{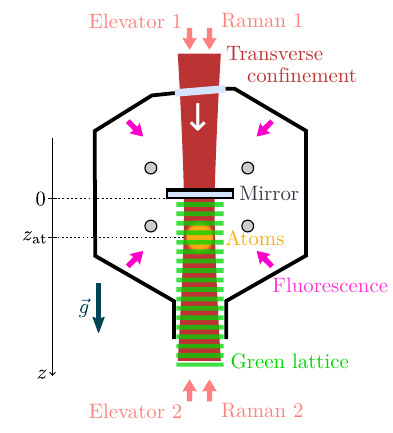}
  \caption{Scheme of the upper chamber setup, with the mirror of interest at its center. Two counter-propagating beams at \qty{780}{nm}, red detuned from the Rb87 D2 line by \qty{250}{GHz}, ensure the transport of the atoms, while two other counter-propagating Raman beams, detuned from the D2 line by \qty{300}{GHz}, are used to drive the interferometer. The Raman 1 beam is actually retroreflected on a Raman mirror, not depicted here, located above the mirror of interest~\cite{pelle_state-labeling_2013}. Atoms end up trapped vertically in a lattice resulting from the retroreflection of a green laser beam at $\lambda_l = \qty{532}{nm}$ onto the surface of the mirror, and transversally with a progressive IR beam at \qty{1064}{nm}. Four linear electrodes (displayed as grey circles) are used to produce external electrostatic fields.
  }\label{fig:montage}
\end{figure}

The transport of the atomic cloud to the vicinity of this mirror is performed with
a moving lattice described thoroughly in~\cite{absil_long-range_2023}. Two counter-propagating beams
(Fig. \ref{fig:montage}) are detuned with a controllable frequency
 difference, creating a moving lattice of controllable speed in which the atoms are trapped. This
 setup allows a very fine control on the transport distance $\Delta z$ with a
resolution in the tens of nanometers.

\begin{figure}
  \centering
  \includegraphics{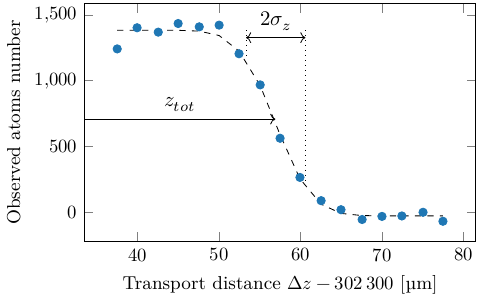} 
  \caption{Number of remaining atoms detected by fluorescence near the mirror surface as a function
  of the transport distance $\Delta z$. We extract the position of the surface $ z_{tot}$ and the
width $\sigma_z$ of the atomic cloud from a fit to the data (dashed line).}\label{fig:calib}
\end{figure}

This method provides us with a precise knowledge of the travelled distance $\Delta z$ from the dipole
trap \cite{absil_long-range_2023}. However, the distance $z_\mathrm{at}$ between the atomic cloud and
the surface is not directly known. This distance depends on
the exact distance $z_{tot} = z_\mathrm{at} + \Delta z$ between the dipole trap and the mirror which fluctuates
depending on beam alignments and thermal variations. 
To calibrate this distance $z_{tot}$ we use the mirror surface directly as a position reference by
transporting our atomic cloud in its vicinity, similarly to \cite{sorrentino_quantum_2009}. Since
atoms kicked into the surface are lost, the exact position of the surface can be determined by 
measuring the number of remaining atoms, as shown in Fig. \ref{fig:calib}. Assuming a Gaussian
density distribution and fitting the atom number by an $\mathrm{erf}$ function, we extract the distance $z_{tot}$, as well as
the $1/e^2$ size of the cloud. This lets us to adjust the atom-surface distance $z$
with sub-micrometer uncertainty, limited by the fit uncertainty. The distance $z_{tot}$ remains
stable at the \qty{0.1}{\um} level during a few hours but fluctuates by a few micrometers over
days, therefore we perform this distance calibration step twice an hour along the force measurements to track any fluctuation.

As for the vertical size $\sigma_z$ of our atom cloud, our measurements show that it remains
unchanged during transport with a $1/e^2$ size $\sigma_z$ of \qty{3.5}{\um}.

We end up with \num{20000} atoms at most after the transport, in a mixture of \ket{F=1} and \ket{F=2}. Atoms
in \ket{F=2} are depumped in \ket{F=1} before the transfer of the atoms in the green static shallow lattice
used for the force measurement. Being blue-detuned, the light potential confines only
in the vertical direction, so we superimpose (Fig. \ref{fig:montage}) an additional IR laser beam, progressive and red-detuned to transversally confines the atoms at the center of the lattice. After
\qty{500}{ms} trapping time, we are left with at most around 1000 trapped atoms in \ket{F=1, m_F=0} to perform the measurements.

Two counterpropagating Raman lasers, detuned  from resonance by \qty{300}{GHz}, couple the $\ket{W_m, F=1}$ and $\ket{W_{m+\Delta m}, F=2}$ states for every initial well number $m$. These beams both enter the chamber at the bottom. At the top exit of the chamber, one of the two Raman beams (Raman 2) is reflected out by a cube while the other (Raman 1) is transmitted and retroreflected on a Raman mirror. This common implementation of the Raman lasers ensures the stability of the Raman phase difference, whose equiphases are tied to the position of the Raman mirror. Using Ramsey interferometry, one can thus probe the energy difference between the states $\ket{W_m, F=1}$ and $\ket{W_{m+\Delta m}, F=2}$ \cite{pelle_state-labeling_2013}. The two output ports of the interferometer being labeled by their hyperfine levels, their populations are measured with \emph{in
situ} state selective fluorescence imaging, using crossing pairs of retroreflected laser beams with an intensity
of \qty{8.5}{mW/cm^2} slightly red-detuned from resonance to enhance the fluorescence collection time, up to
\qty{1}{\ms}. An imaging system with a numerical aperture of 0.32 collects about 40 photons per
atoms onto an Electron Multiplying CCD (EMCCD) camera.

To measure the Bloch frequency, we perform a Ramsey interferometer, with two Raman $\pi/2$ pulses separated by the interrogation time
$T$.
The output phase $\Phi$ for this interferometer is given by $
  \Phi = 2 \pi (\nu_R - \nu_{HFS} - \Delta m \,\nu_B)\, T$
where $\nu_R$ is the frequency difference between the two Raman beams and  $\nu_{HFS}$ is the
hyperfine splitting frequency. By tuning the Raman frequency difference to keep the output phase
null, we get $\
\nu_R = \nu_{HFS} + \Delta m \,\nu_B$.  

To remove the dependence on the hyperfine frequency, which can slowly fluctuate with laser light shifts, we alternately perform measurements for $\pm \Delta m$. The difference between the two measurements gives the Bloch frequency free from the hyperfine frequency $\nu_B = (\nu_R^{+\Delta m} - \nu_R^{-\Delta m})/2 \Delta m$.

Note that all six vertical beams (Fig. \ref{fig:montage}) are overlapped, in order to avoid
parasitic couplings to transverse states \cite{tackmann_raman_2011}. The uncertainty in the overall vertical alignment has been reduced down to 1.5 mrad using a liquid mirror as a reference, while all beams are superimposed on one another to better than 0.1 mrad. 

We now quickly discuss the limitations of the sensitivity of the force measurement. 
The contrast of the interferometer is typically of \qty{40}{\percent} for a coherence  time of about
\qty{150}{ms}. It is mainly limited by inhomogeneities in the light shift. Additionally, at less than \qty{20}{\um} of the mirror, inhomogeneities of the atom-surface force decrease further the interferometer contrast.

As for the measurement noise, we identified three main contributions. The first is Raman phase
noise due to vibrations of the mirror of interest relatively to the Raman retroreflecting mirror.  
The support of the mirror of interest is attached to the end of a 3-axis in-vacuum manipulator, clamped
to the overall structure with sorbothane pads used for isolation in order to suppress undesired vibration induced tunneling between neighbouring wells of the lattice \cite{Ivanov2008}. Still, residual vibrations are estimated to induce fluctuations on $\nu_B$ at the level of \qty{480}{mHz} rms per shot. 
By implementing an optical Michelson interferometer that combines the incoming and retroreflected lattice laser beams, we are able to track vibrations of the mirror of interest relatively to the optical bench. By correcting \emph{a posteriori} the Bloch frequency measurement from the effect of measured
vibrations, similar to \cite{legouet2008}, we reduce the contributions of vibration noise by about 30\%, down to the level of \qty{300}{mHz} per shot.

The second noise source is the quantum projection noise, corresponding to \qty{200}{mHz} for the typical atom number we had in the measurements, of the order of 500. The third is a background detection noise, equivalent to 20 atoms, which adds
\qty{190}{mHz}. These three contributions sum up (quadratically) to \qty{410}{mHz}, which is in agreement with the measured noise of \qty{430}{mHz}. 
At the end, since we use $\Delta m = 6$, we reach a sensitivity on the measurement of the Bloch
frequency of \qty{71}{mHz}/shot.

Given our cycle time of \qty{3.6}{s}, this corresponds to a sensitivity of \qty{3.4e-28}{N} at
\qty{1}{s}. For an averaging time of \qty{5.5}{h},
the noise averages down as white noise and a long-term stability of 4 quectonewtons is reached.
This lies above the best sensitivity of \qty{7e-30}{N} at \qty{1}{s} reported both with this method \cite{alauze_trapped_2018} and an alternative one based on the amplitude modulation of the lattice \cite{Tarallo2014}, but far from any surface. But it outperforms any other local force measurements, such as based on monitoring the dynamics of trapped atoms \cite{guo_QuantumPrecisionMeasurement_2022} or ions \cite{ivanov_HighprecisionForceSensing_2016,
gilmore_QuantumenhancedSensingDisplacements_2021, harlander_TrappedionProbingLightinduced_2010}, or of microscopic objects \cite{ranjit_ZeptonewtonForceSensing_2016}. It is to the best of our knowledge state-of-the-art for surface force measurements.

\begin{figure}
  \includegraphics{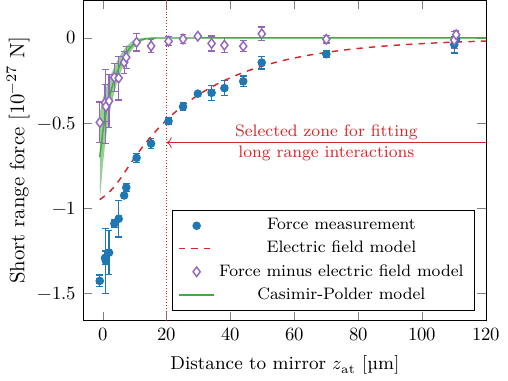}
  \caption{Force measurements close to the mirror surface. Blue circles: force measurements. Dashed
    red: fit to these above \qty{20}{\um} by a model of the interactions with a Gaussian
  distribution of electric dipoles. Purple diamonds: difference between this model and the measurements. Green line: expected Casimir-Polder force for our setup. The error bars denote the 1-sigma statistical errors.}\label{fig:force}
\end{figure}

This high sensitivity associated with the control of the distance to the mirror allows us to perform sensitive force measurements in the vicinity of the mirror. By performing interleaved differential measurements close to the mirror and further away at a distance of \qty{300}{\um}, we suppress the gravity force and any long-range bias on the force, such a parasitic vertical dipole force from the IR transverse confinement beam. This dipole force has been evaluated by varying the IR power as an offset of the order of \qty{3e-29}{N} with no resolved dependence on the distance to the mirror.
These differential force measurements, displayed Fig. \ref{fig:force}, exhibit a clearly attractive
behaviour. As a reference, the effect expected from the Casimir-Polder force is also displayed as a
green curve. It was calculated using the energy shifts of the Wannier Stark states due to the
Casimir-Polder interaction, as computed in \cite{Maury2016,maury_effet_2016} for our exact
experimental configuration. From a simulation of the fringe patterns averaged over the finite size of the cloud, we extract the frequency shift of the central fringe, and the corresponding expected force.

The observed force, which has a longer range and greater magnitude, cannot be explained by the Casimir-Polder effect alone.
Another interaction is at play, likely due to electrostatic interactions between our atomic sample and electric dipoles by atoms adsorbed on the dielectric surface, as already put into evidence in
\cite{mcguirk_alkali-metal_2004}. 

Rb atoms adsorbed on neutral surfaces create dipole electric moments whose value has been
evaluated to $\mu = 3.2$ D in \cite{obrecht_measuring_2007} for a silica substrate. An atom brought
near the surface thus experiences an attractive potential $U_\mathrm{elec} = \frac{\alpha_0} 2 |\vec E|^2$ where $\alpha_0$ is the static polarizability of the atom ground state and $\vec E$ is the electrostatic field produced by these moments.
So a force $\vec F_\mathrm{elec} = -\frac{\alpha_0}2\ \vec\nabla |\vec E|^2$ is exerted on our atoms, to the vertical component of which our sensor is sensitive.

We observed that we were able to reduce by half the force measured for distances above \qty{20}{\um}, where the Casimir-Polder effect is weak, by simply turning off the experiment for a month. It returned to a new steady state after a week of operation. We interpreted this as a first evidence supporting the idea of stray electric fields from adsorbed atoms, which slowly desorb or diffuse, as already demonstrated in \cite{obrecht_measuring_2007}.

Furthermore, by applying uniform vertical external electric fields $E_\mathrm{ext}^z$ using electrodes surrounding the mirror as shown Fig. \ref{fig:montage}, we have been able to get a direct evidence of these electric fields. 
This amplifies the force $F^z_\mathrm{elec}$ by $-\alpha_0 E_\mathrm{ext}^z {\textnormal d}_z{E^z}$ and allows for
measuring vertical gradients of the parasitic electric field ${\textnormal d}_z{E^z}$.  First
measurements show a measured vertical electric gradient of the order of \qty{2.0 (1)e7}{V/m^2} at a
distance of \qty{20}{\um}, in marginal agreement with the gradient expected from a model
described below (\qty{3.3(4)e7}{V/m^2}).

To model these parasitic electric fields, we consider a Gaussian distribution of $N$ dipole moments over an rms radius $\sigma_r$: $n(r) = N /( {2 \pi
{\sigma_r}^2}) \allowbreak \exp(- r^2 / (2 {\sigma_r}^2))$.
The resulting electrostatic force along the vertical direction depends only on the two parameters
$N$ and $\sigma_r$ which we estimate by fitting our force measurements in the range $z >
\qty{20}{\um} $ where the Casimir-Polder force is negligible, see Fig. \ref{fig:force}. 
This gives us an estimate for $\sigma_r$ around $\qty{90}{\um}$ and $N$ around $\num{2.0e10}$, corresponding to a radius twice as large as the radial width of the atomic cloud at the end of the transport, and a number of adsorbed atoms compatible with a continuous launch of atoms over several weeks of the running experiment, during which most of the atoms are not stopped at the end of the transport.

Remarkably, by using this estimated bias electrostatic force to correct our force measurements over the full range, we obtain good agreement between these corrected points, depicted with diamond points in Fig. \ref{fig:force}, and the expected Casimir-Polder force in the range
$z < \qty{20}{\um} $.

In conclusion, we demonstrated a state-of-the art short-range force sensor using atom interferometry inside an optical lattice. In the regime of distance explored, the electrostatic force from adsorbed atoms predominates, but we achieved an evidence of Casimir-Polder force from a mirror in the micrometer range. 
A thorough characterization of stray electric fields would allow for a better comparison between
expected and measured Casimir Polder forces, opening the way for tests of gravity at short range.
The removal, or at least the reduction of the number of adsorbed atoms, thanks to the heating of the
surface \cite{obrecht_measuring_2007} would be an asset for such studies. 
We expect to get better spatial resolutions in the future, starting with even smaller clouds or implementing spatial selection.
This selection could be realized by lifting the degeneracy between Wannier-Stark transitions, for
instance by using magnetic field gradients \cite{Schrader2004}, or by shaping the trapping
potential as a superlattice \cite{pereira2009}. The resulting loss in the number of atoms could be mitigated by improving the efficiency of the atom transport to increase the number of atoms. This would allow for even more sensitive and spatially resolved measurements of atom-surface interactions.

\begin{acknowledgements}
We thank Xiaobing Deng for early contributions, and Axel Maury and his colleagues for providing us with the calculations of the energy levels in our system. This research has been carried out in the frame of the QuantERA project TAIOL, funded by the European Union’s Horizon 2020 Research and Innovation Programme and the Agence Nationale de la Recherche (ANR-18-QUAN-0015-01). 
\end{acknowledgements}
Y.B. and L.A. contributed equally to this work.

\bibliography{biblio}

\end{document}